\documentclass[aps,prl,superscriptaddress,twocolumn,showpacs]{revtex4}

\renewcommand*{\[}{\begin{equation}}

\renewcommand*{\]}{\end{equation}}

\def\PRA{{Phys.~Rev.~A} }

\def\JPB{{J.~Phys.~B} }
\def\PRL{{Phys.~Rev.~Lett.} }

\newcommand{\myscaleboxa}[1]{\scalebox{0.30}[0.30]{#1}}
\newcommand{\myscaleboxb}[1]{\scalebox{0.44}[0.44]{#1}}

\usepackage{epsfig}

\usepackage{hyperref}

\begin{document}

\title{Analysis of two-dimensional high-energy photoelectron momentum distributions in single ionization of atoms by intense laser pulses}

\author{Zhangjin Chen}

\affiliation{J. R. Macdonald Laboratory, Physics Department, Kansas
State University, Manhattan, Kansas 66506-2604, USA}

\author{Toru Morishita}

\affiliation{Department of Applied Physics and Chemistry, The
University of Electro-Communications, 1-5-1 Chofu-ga-oka, Chofu-shi,
Tokyo 182-8585, Japan}

\author{Anh-Thu Le}

\affiliation{J. R. Macdonald Laboratory, Physics Department, Kansas
State University, Manhattan, Kansas 66506-2604, USA}

\author{C. D. Lin }

\affiliation{J. R. Macdonald Laboratory, Physics Department, Kansas
State University, Manhattan, Kansas 66506-2604, USA}

\date{\today}

\begin{abstract}

We analyzed the two-dimensional (2D) electron momentum
distributions of high-energy photoelectrons of atoms in an intense
laser field using the second-order strong field approximation
(SFA2). The SFA2 accounts for the  rescattering of the returning
electron with the target ion to first order and its validity is
established by comparing with results obtained by solving the
time-dependent Schr\"{o}dinger equation (TDSE) for short pulses.
By analyzing the SFA2 theory, we confirmed that the yield along
the back rescattered ridge (BRR) in the 2D momentum spectra can be
interpreted as due to the elastic scattering in the backward
directions by the returning electron wave packet. The
characteristics of the extracted electron wave packets for
different laser parameters are analyzed, including their
dependence on the laser intensity and pulse duration. For long
pulses we also studied the wave packets from the first and the
later returns.

\end{abstract}

\pacs{32.80.Rm, 32.80.Fb, 42.50.Hz}

\maketitle

\section{Introduction}
Much of our knowledge of strong field, multiphoton ionization
processes comes from the study of above-threshold ionization (ATI).
In this process an atom absorbs more photons than the minimum number
required for the ejection of an electron. By measuring the energy of
the photoelectrons, the characteristic ATI spectra are peaks
separated by the photon energy, with the peak positions shifted by
the ponderomotive potential. For long pulses, additional
substructures due to Freeman resonances \cite{Freeman} appear. For
shorter pulses, of durations of the order of 10-20 fs or less,
calculations have shown that the major ATI peaks are accompanied by
subpeaks which have been attributed to the rapidly changing
ponderomotive potential \cite{marlene1,bardsley}.

Experimentally more   information on ATI electrons can be
determined by measuring the angular distributions or the 2D
momentum distributions. For example, the nature of Freeman
resonances are associated with individual Rydberg states by the
number of lobes in their angular distributions \cite{helm03}. For
short pulses in the tunneling ionization regime, recent
experiments \cite{ullrich,cocke} have shown ubiquitous fan-like
structures in the 2D momentum spectra for low-energy electrons.
These fan-like structures are now also well-understood and they
are due to the long-range Coulomb interaction between the tunnel
ionized electron and the target ion it has left behind
\cite{chen06,Arbo}. These fan-like structures do survive the
integration over the laser focus volume, as shown in recent
comparison between theoretical calculations with experiments
\cite{toru}.

High-energy photoelectrons have been measured previously for atomic
targets, both in energy and angular distributions, using longer
pulses at lower intensities. The energy spectra above $4 U_p$, where
$U_p$ is the ponderomotive energy, have been found to vary rapidly
with small changes in laser intensities \cite{hertlein,paulus01}
when laser pulse durations are larger than 20 to 30 fs. Different
theoretical models have been used to interpret these phenomena, but
the conclusions are still tentative so far
\cite{muller,maquet,becker,tony}. While single ionization of an atom
in an intense laser field has been calculated accurately by solving
the time-dependent Schr\"{o}dinger equation (TDSE) directly in the
past two decades, the photoelectrons in the high-energy region,
especially their momentum distributions, have not been investigated.
This is not surprising since the photoelectron yield drops rapidly
with the electron's energy. On the other hand, it is known that
rescattering plays a major role in many laser-atom interaction
phenomena, including the generation of high-energy ATI electrons.
Understanding the nature of these high-energy photoelectrons,
especially their 2D momentum distributions, may help to shed new
light on the rescattering process itself.

Recently we have initiated a careful study on the 2D momentum
distributions of high-energy photoelectrons by directly solving the
TDSE of a one-electron atom in a short laser pulse. By focusing on
photoelectrons that have been backscattered we have been able to
extract the elastic scattering cross sections by free electrons from
the calculated laser-generated photoelectron spectra. For molecular
targets this has the important significance that it offers the
possibility of using infrared lasers for dynamic chemical imaging
with temporal resolution of a few femtoseconds \cite{torunew}. For
short laser pulses, by studying the dependence of the 2D momentum
spectra on the carrier-envelope phase, it also offers the
possibility of directly characterizing the electric fields of
few-cycle pulses.

To obtain 2D momentum electron spectra accurately in the
high-energy region, say, up to about 10 $U_p$, calculations have
to be done very carefully. For pulses as short as a few
femtoseconds and intensities with Keldysh parameters $\gamma$
\cite{Keldysh} close to one, we have been able to obtain accurate
2D momentum spectra from solving TDSE directly. Here
$\gamma=\sqrt{I_p/(2U_p)}$, with $I_p$ the binding energy of the
electron, and $U_p=I_{\text max}/(4\omega^2)$, where $I_{\text
max}$ is the peak intensity and $\omega$ is the carrier frequency,
of the laser, respectively. (Atomic units are used throughout
unless indicated otherwise.) To extend such TDSE calculations to
high laser intensities or long durations is computationally more
challenging. We thus seek to examine the predictions based on the
second-order strong-field approximation (SFA2)
\cite{lew2,milosevic03OP,milosevic03PRA,milosevic04PRA,milosevic05,faiser}
where the rescattering of the returning electron with the target
ion is included. To establish the validity of SFA2, we first
compare its results with those from solving the TDSE for short
pulses. The SFA2 is then used to analyze high-energy photoelectron
momentum distributions for various laser parameters. From such
analysis, we show that electrons on the so-called back rescattered
ridge (BRR) can be identified. The electron yield along the BRR
can be used to extract elastic scattering cross sections of free
electrons by target ions, as well as the wave packet of the
returning electrons. Using SFA2 allows us to analyze the electron
wave packets for longer pulses and higher intensities.

In the next Section we first summarize the first- and second-order
amplitudes of the strong field approximation. The results from
using this theory for the prediction of the 2D momentum spectra
are shown in Section III. The conclusion is given in Section IV.

\section{Theoretical models}

To study atoms in an intense laser field, there are only two general
theoretical tools at present. One is to numerically integrate the
TDSE and the other is based on the SFA. Both approaches have their
limitations. For TDSE, to obtain high precision in small quantities
such as the momentum distributions of high-energy electrons poses a
computational challenge. This challenge is less severe for short
pulses where the electron is confined to a smaller box such that
accurate solution of the TDSE is possible. For longer pulses and
higher intensities, high accuracy becomes harder to achieve. One has
to check carefully the possible effect of reflection from the box
boundaries as well as the convergence of the basis set used. While
reflection can be avoided by matching the solution inside the box to
the outside region where it is expanded in terms of Volkov states
\cite{tong},  such an approach requires large basis set in both the
inside and the outside regions, and thus the method is also very
computationally intensive. Thus in the TDSE calculations we limit
ourselves only to short pulses. The numerical method we used for
solving TDSE has been presented in our previous works \cite{chen06,
toru}. In this paper we only present the details of the SFA2 used
here.

An exact expression for the probability amplitude of detecting an
ATI electron with momentum $\textbf{p}$ can be written formally as
\begin{eqnarray}
\label{exact}f({\textbf{p}})=-i\lim_{t\rightarrow\infty}
\int_{-t}^{t}dt^{\prime}\left\langle\Psi_{\textbf{p}}(t)\left|U(t,t^{\prime})H_i(t^{\prime})\right|\Psi_{0}(t^{\prime})\right\rangle.
\end{eqnarray}
Here $U(t,t^{\prime})$ is the time-evolution operator of the
complete Hamiltonian
\begin{eqnarray}
H(t)=H_a+H_i(t)
\end{eqnarray}
where
\begin{eqnarray}
H_a=-\frac{1}{2}\nabla^2+V(\textbf{r})
\end{eqnarray}
is the atomic Hamiltonian, and
\begin{eqnarray}
H_i(t)=\textbf{r}\cdot\textbf{E}(t)
\end{eqnarray}
is the laser-electron interaction in the length gauge and the
dipole approximation. The linearly polarized electric field
$\textbf{E}(t)$ of the laser pulse along the $z$ axis is given by
\begin{eqnarray}
\textbf{E}(t)=E_{0}a(t)\cos(\omega t+\phi)\hat{z}
\end{eqnarray}
where $\phi$ is the carrier-envelope phase. The envelope function
$a(t)$ is chosen to be
\begin{eqnarray}
\label{envelope}a(t)=\cos^{2}\left(\frac{\pi t}{T}\right)
\end{eqnarray}
for the time interval ($-T/2,T/2$), and zero elsewhere. In this
paper, $T$ is defined as the (full) duration of the laser pulse
which is 2.75 times of the FWHM (full width at half maximum) and the
carrier-envelope phase $\phi$ is set as zero. The functions
$\Psi_{\textbf{p}}(t)$ and $\Psi_{0}(t)$ are scattering state with
asymptotic momentum $\textbf{p}$ and the ground state, respectively,
of the atomic Hamiltonian $H_a$. The time-evolution operator
$U(t,t^{\prime})$ satisfies the Dyson equation
\begin{eqnarray}
U(t,t^{\prime})=U_{\text{F}}(t,t^{\prime})-i\int_{t^\prime}^{t}dt^{\prime\prime}
U_{\text{F}}(t,t^{\prime\prime})VU(t^{\prime\prime},t^{\prime})
\end{eqnarray}
where $U_{\text{F}}(t,t^{\prime})$ is the time-evolution operator
for the Hamiltonian of a free electron in the laser field, which
is
\begin{eqnarray}
H_{\text{F}}=-\frac{1}{2}\nabla^2+\textbf{r}\cdot\textbf{E}.
\end{eqnarray}
The eigenstates of  $H_{\text{F}}(t)$ are the Volkov states
\begin{eqnarray}
\left|\chi_{\textbf{p}}(t)\right\rangle=\left|\textbf{p}+\textbf{A}(t)\right\rangle\exp[-iS_{\textbf{p}}(t)]
\end{eqnarray}
with the action
\begin{eqnarray}
S_{\textbf{p}}(t)=\frac{1}{2}\int_{-\infty}^t
dt^{\prime}\left[\textbf{p}+\textbf{A}^{\prime}(t)\right]^2.
\end{eqnarray}
The vector potential of the laser field $\textbf{E}(t)$ is denoted
by $\textbf{A}(t)$, and $\left|\textbf{k}\right\rangle$ is a plane
wave state
\begin{eqnarray}
\langle \textbf{r}|\textbf{k}\rangle=\frac{1}{(2 \pi)^{3/2}}\exp
\left(i\textbf{k} \cdot \textbf{r} \right).
\end{eqnarray}
The Volkov time-evolution operator is
\begin{eqnarray}
U_{\text{F}}(t,t^{\prime})=\int
d\textbf{k}\left|\chi_{\textbf{k}}(t)\right\rangle\left\langle\chi_{\textbf{k}}(t^{\prime})\right|.
\end{eqnarray}

By approximating $U(t^{\prime\prime},t^{\prime})$ on the right-hand
side of Eq. (7) by $U_F(t^{\prime\prime},t^{\prime})$, and
$\left\langle\Psi_{\textbf{p}}(t)\right|$ in Eq. (\ref{exact}) by
$\left\langle \chi_{\textbf{k}}(t)\right|$, the ionization amplitude
may be expressed as
\begin{eqnarray}
\label{full_SFA}f =f^{(1)}+f^{(2)}
\end{eqnarray}
where the first term
\begin{eqnarray}
\label{1st-order}f^{(1)}=-i
\int_{-\infty}^{\infty}dt\left\langle\chi_{\textbf{p}}(t)\left|H_i(t)\right|\Psi_{0}(t)\right\rangle
\end{eqnarray}
corresponds to the standard SFA. This term will be called SFA1 for
the present purpose. The second term is the SFA2,
\begin{eqnarray}
\label{2nd-order}f^{(2)}&=&-\int_{-\infty}^{\infty}dt\int_{-\infty}^{t}dt^{\prime}\int
d\textbf{k}\left\langle\chi_{\textbf{p}}(t)\left|V\right|\chi_{\textbf{k}}(t)\right\rangle
\nonumber \\ && \times
\left\langle\chi_{\textbf{k}}(t^{\prime})\left|H_i(t^{\prime})\right|\Psi_{0}(t^{\prime})\right\rangle
\end{eqnarray}
which accounts for the first-order correction by the atomic
potential. This expression can be easily understood by reading it
from the right side. The electron is first ionized at time
$t^{\prime}$ by the laser field. It then propagates in the laser
field from $t^{\prime}$ to $t$ where it is rescattered by the
atomic potential $V$ into a state with momentum $\textbf{p}$. Note
that in this approximation, the interaction of the electron with
the atomic potential is treated up to the first order only.

The evaluation of the matrix elements are illustrated below for
the hydrogen-like atoms where the ground state wavefunction takes
the form
\begin{eqnarray}
\Psi_{\text{1s}}(\textbf{r})=2Z^{3/2}\exp(-Zr)Y_{00}(\hat
{\textbf{r}})
\end{eqnarray}
where $Z$ is the charge of nucleus. The rescattering potential
$V(\textbf{r})$ is a pure Coulomb potential
\begin{eqnarray}
\label{pot}V(\textbf{r})=-\frac{Z}{r}.
\end{eqnarray}
In the numerical calculations, the Coulomb potential is replaced
by Yukawa potential with a damping parameter $\alpha$ to avoid the
singularity in the integrand in $f^{(2)}$,
\begin{eqnarray}
\label{short_pot}\tilde{V}(\textbf{r})=V(\textbf{r})e^{-\alpha r}.
\end{eqnarray}
This introduces a very weak dependence of the magnitude of $f^{(2)}$
on the value of $\alpha$, but not the shape.  Here, we chose
$\alpha=1.0$.

After performing integration  over space coordinates analytically,
the amplitudes $f^{(1)}$ and $f^{(2)}$ become
\begin{eqnarray}
\label{1st-order_ana}f^{(1)}&=&-\frac{8\sqrt{2}}{\pi}Z^{5/2}
\int_{-\infty}^{\infty}dt \exp\left[i
S_{\textbf{p}}(t)\right]\exp(iI_pt)\nonumber
\\ &&\times
\frac{\textbf{E}(t)\cdot[\textbf{p}+\textbf{A}(t)]}
{\left\{Z^2+[\textbf{p}+\textbf{A}(t)]^2\right\}^3}
\end{eqnarray}
and
\begin{eqnarray}
\label{2nd-order_ana}f^{(2)}&=&-i\frac{4\sqrt{2}}{\pi^3}Z^{5/2}
\int_{-\infty}^{\infty}dt\int_{-\infty}^{t}dt^{\prime}\int
d\textbf{k} \nonumber \\ &&\times
\exp\left\{-i[S_{\textbf{k}}(t)-S_{\textbf{p}}(t)]\right\}
\exp[iS_{\textbf{k}}(t^{\prime})]\exp(iI_p t^{\prime}) \nonumber \\
&&\times \frac{1}{\alpha^2+(\textbf{k}-\textbf{p})^2}
\frac{\textbf{E}(t^{\prime})\cdot[\textbf{k}+\textbf{A}(t^{\prime})]}{\left\{Z^2+[\textbf{k}+\textbf{A}(t^{\prime})]^2\right\}^3}
\end{eqnarray}
respectively. The evaluation of the first-order amplitude
(\ref{1st-order_ana}) is straightforward. The second-order amplitude
(\ref{2nd-order_ana}) consists of fivefold integration. We used
saddle point approximation for the integration with respect to
$\textbf{k}$, as proposed in Lewenstein \emph{et al} \cite{lew2}, to
reduce it to a twofold one. The saddle point is calculated with
respect to  the quasiclassical action only, so that
\begin{eqnarray}
\label{sad_k}\textbf{k}_S(t,t^{\prime})=-\frac{1}{t-t^{\prime}}
\int_{t^{\prime}} ^t dt^{\prime\prime}\textbf{A}(t^{\prime\prime}).
\end{eqnarray}
The result of saddle point integration for the fivefold integral
(\ref{2nd-order_ana}) is obtained by setting
\begin{eqnarray}
\textbf{k}=\textbf{k}_S(t,t^{\prime}),
\end{eqnarray}
and substituting
\begin{eqnarray}
\label{sad_k2}\int d\textbf{k} \rightarrow
\left[\frac{2\pi}{\epsilon+i(t-t^{\prime})}\right]^{3/2}
\end{eqnarray}
for the integration over $\textbf{k}$. Here, $\epsilon$ is an
arbitrary small parameter introduced to smooth out the singularity
in (\ref{sad_k2}). Its value is taken to be of the order of 0.1.

The momentum distribution of the emission of an electron of energy
$E=p^2/2$ in the direction of $\hat{\textbf{p}}$ is given by
\begin{eqnarray}
\frac{\partial ^2 P}{\partial E \partial
\label{probability}\hat{\textbf{p}}}=|f|^2.
\end{eqnarray}
The form of $f$ in (\ref{full_SFA}) allows us to identify the
contribution from each individual terms SFA1 and SFA2,
respectively.

For a linearly polarized laser field, the system has cylindrical
symmetry for the cases considered here. As a result, the
two-dimensional momentum distribution is defined by
\begin{eqnarray}
\label{2D_momentum}\frac{\partial ^2 P}{\partial E \partial
\theta}=\frac{\partial ^2 P}{\partial E \partial \hat{\textbf{p}}}
2\pi p \sin\theta
\end{eqnarray}
where integration over the azimuthal angle $\varphi$ has been
carried out and $\theta$ is the angle between the polarization
axis of the laser field and the direction of the ejected
photoelectron. By integrating over $\theta$ in equation
(\ref{2D_momentum}), we obtain the energy spectra
\begin{eqnarray}
\label{energy}\frac{\partial P}{\partial E}=\int \frac{\partial ^2
P}{\partial E \partial \theta} d\theta.
\end{eqnarray}

\section{Results and discussion}

\subsection{Validity of the SFA2}
First we establish the validity and the limitation of the SFA2. In
Fig. 1(a) we show the total ionization probability vs electron
energy for a hydrogen atom ionized by a 5-cycle (full duration)
laser pulse with the wavelength of 800 nm and at the peak
intensity of $1.0\times 10^{14}$ W/cm$^{2}$. For this case, the
ponderomotive energy is 6 eV. Within the perturbation approach we
note that the yield is dominated by SFA1 for low energy electrons.
For energies greater than about $4 U_p$, SFA2 becomes dominant.
Interference between SFA1 and SFA2 is important only in a small
energy region. Note that the actual probability obtained from SFA1
severely underestimates the total ionization yield as  obtained by
the TDSE. This underestimate is present also in SFA2 since the
same matrix element for the initial ionization of the atom is also
used in SFA2. Here we are interested in the high-energy region, we
thus renormalize the electron energy spectra to the TDSE results
at higher energies. Note that after normalization, see Fig. 1(b),
the electron energy spectra from the two theories are in good
agreement for energies above $5 U_p$. This comparison also proves
that rescattering plays a major role in the high-energy ATI
electrons.

\begin{figure}
\mbox{\rotatebox{270}{\myscaleboxa{
\includegraphics{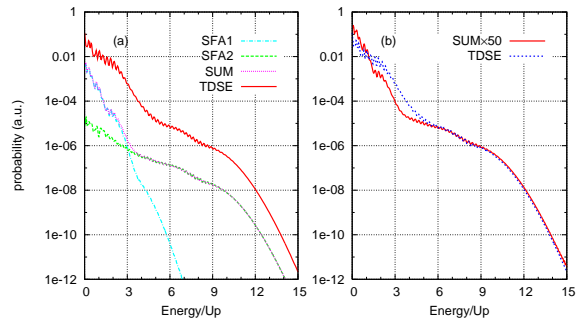}}}}
\caption{(Color online) Electron energy spectra for atomic hydrogen
by a 5-cycle laser pulse with the wavelength of 800 nm at the peak
intensity of $1.0\times 10^{14}$ W/cm$^{2}$.(a) From TDSE, first-
(SFA1) and 2nd-order (SFA2) theory and the coherent sum of SFA1 and
SFA2; (b) the sum of SFA1 and SFA2 but renormalized to the
high-energy part of TDSE. Electron energy is in units of $U_p$, the
ponderomotive energy. }
\end{figure}

We next consider the 2D electron momentum spectra. Previously
theoretical calculations tend to focus on the angular distributions
at specific angles, especially along the laser polarization
direction \cite{Dionissopoulou97, bauer06}. Our goal, instead, is to
examine the global 2D momentum spectra in the high-energy region and
whether the spectra can be described by the SFA2. In Fig. 2 the 2D
momentum spectra calculated from SFA2 and from TDSE are shown. The
horizontal and the vertical axes are the electron momenta parallel
and perpendicular to the laser polarization, respectively, on a
plane containing the polarization axis. Since the electron yield
drops very rapidly with energy (see Fig. 1), to make the high energy
part visible, the 2D momentum distributions in each frame have been
renormalized such that the total ionization yield at each electron
energy is the same. At first glance, clearly the two frames look
very similar in the high-momentum region. (The difference in the
small momentum part is not important since SFA1 is dominant there.)
Similar agreement has also been observed for lasers of different
intensities and wavelengths.

The striking features of Fig. 2 are the two half circles on each
side of the origin. The center of each circle is shifted along the
$p_{\parallel}(=p_z)$ axis. We call these circular rings back
rescattered ridge (BRR), representing electrons that have been
rescattered into the backward directions by the target ion. Based
on the results from TDSE calculations, we have confirmed
\cite{torunew} that the electron yields on the BRR can be
interpreted as elastic scattering of the returning electrons by
the target ion potential. Since the main features  of the BRR are
also reproduced in the SFA2 we seek to provide theoretical grounds
for this interpretation.

\begin{figure}
\mbox{{\myscaleboxa{
\includegraphics{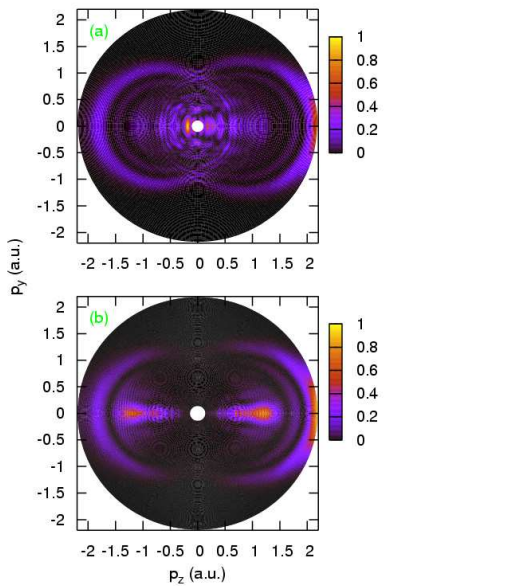}}}}
\caption{(Color online) Photoelectron 2D momentum distributions
parallel ($p_\parallel=p_z$) and perpendicular ($p_y$) to the laser
polarization direction for atomic hydrogen by a 5-cycle laser pulse
with the wavelength of 800 nm at the peak intensity of $1.0\times
10^{14}$ W/cm$^{2}$. (a) SFA2 and (b) TDSE.}
\end{figure}

Evidence of these BRR electrons had been seen observed previously.
Using 50-ps, 1.05-${\mu}$m linearly polarized laser pulses, Yang
\emph{et al}. \cite{yang93} observed unexpected narrow lobes in
the angular distributions at approximately 45$^{\circ}$ off the
polarization axis in the high order ATI spectrum around $9U_p$ in
single ionization of Xe and Kr atoms. The origin of the sidelobes
has been analyzed by Paulus \emph{et al}. \cite{Paulus_JPB94} in a
two-step classical model and by Lewenstein \emph{et al}.
\cite{lew2} using a quasiclassical analysis. Similar theoretical
analysis of these narrow sidelobes has been made by Dionissopoulou
\emph{et al}. \cite{Dionissopoulou97}. However, direct connection
of these narrow sidelobes with the backscattering of the returning
electrons has not been established at the quantitative level so
far.

\subsection{Analysis of the high-energy 2D momentum spectra}

To identify that rescattering is responsible for the high-energy
BRR electrons, in Fig. 3 we show the renormalized 2D momentum
spectra for 2-cycle and 3-cycle 800 nm pulses at peak intensity of
$1.0\times10^{14}$ W/cm$^2$, with the corresponding electric
fields and vector potentials plotted on the right side. The data
in the 2D momentum spectra are integrated over the azimuthal angle
around the polarization vector, as   in (\ref{2D_momentum}). This
presentation forces the distribution to go to zero on the
$p_\parallel$ axis.

Consider the 2-cycle pulse, from Fig. 3(b) the electric field $E(t)$
reaches local maximum at time $t=-0.4\tau$, 0, and $0.4\tau$, where
$\tau=2\pi/\omega$ is the period of the laser. Based on classical
theory, electrons that are "born" near the peak of the electric
field (more precisely, at the phase angle of 17$^{\circ}$ after the
peak) will be driven back to the target ion about three quarters of
a cycle later. Consider an electron that was born at the time $t_b$
near $-0.4\tau$ where the electric field strength is maximum. We
will choose the convention that positive $z$ is the right side and
negative $z$ the left side. The released electron is in a negative
electric field so it will first moves to the right side. After the
electric field changes to positive, the electron will be decelerated
and may be driven back toward the target. Simple classical
calculation shows that it will return to the origin at time $t_r$
near $0.25\tau$ where the electric field returns to zero. If this
electron is scattered in the forward direction, it will be
decelerated by the subsequent negative electric field for $t>t_r$
and ending up with low energy. On the other hand, if the electron is
backscattered elastically, it will be further accelerated by the
laser's electric field and ending up with high energy. This effect
on the electron momentum is obtained by adding the instantaneous
vector potential ${\bf A}_r\equiv{\bf A}(t_r)$ to the canonical
momentum ${\bf p}$, analogous to the "streaking" of an electron
generated by an X-ray attosecond pulse in the presence of a
femtosecond IR pulse \cite{auger}. The shift of the center of the
BRR is a measure of the magnitude of ${\bf A}_r$, and the radius of
the circle is related to the maximum electron's returning energy
$3.17U_p$. In Fig.~3(b), electrons that are born near $t=0$, would
return at about $t=0.75\tau$ where the vector potential and the
electric field are near zero, thus no high-energy electrons would
emerge. This explains why there is no BRR on the left side of Fig.
3(a) for the 2-cycle pulse.

\begin{figure}
\mbox{\rotatebox{270}{\myscaleboxa{
\includegraphics{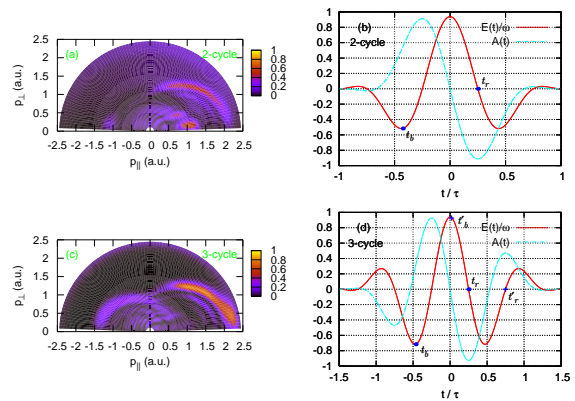}}}}
\caption{(Color online) Photoelectron 2D momentum distributions
parallel ($p_\parallel=p_z$) and perpendicular
($p_{\perp}=\sqrt{p_x^2+p_y^2}$) to the laser polarization direction
for atomic hydrogen by laser pulses at the peak intensity of
$1.0\times 10^{14}$ W/cm$^{2}$ with the wavelength of 800 nm for
durations of 2 cycles and 3 cycles, respectively. The corresponding
laser fields and vector potentials are shown on the right for the
analysis of the rescattering mechanism.}
\end{figure}

Consider next the 3-cycle pulse, the right BRR is due to
ionization occurs near $t_b$ and returns near $t_r$ as before. The
BRR on the left is due to ionization near $t^{\prime}_b$ and
returns near $t^{\prime}_r$. Since the vector potential at
$t^{\prime}_r$ is smaller than that at $t_r$, the shift of the
center and the radius of the circle of the left BRR are smaller.

One advantage of the SFA2 is that it allows us to identify the born
time and the returning time directly.  To analyze the right-side BRR
in Fig. 3(c), we set "window functions" such that the born time is
restricted to $[-0.5\tau,-0.4\tau]$ for $t^{\prime}$ and the return
time interval $[0,0.75\tau]$ for $t$ in (\ref{2nd-order_ana}), as
shown in Fig. 4(b). The resulting 2D spectra shown in Fig. 4(a) is
similar to the right side of the 2D spectra in Fig. 3(c), confirming
that the right BRR is from electrons born near $t_b$. Similarly, in
Fig. 4(c), the 2D momentum distributions are presented to identify
the left-side ridge by setting the born time interval $[0,0.1\tau]$
for $t^{\prime}$ and the returning time interval $[0.5\tau,1.5\tau]$
for $t$ in (\ref{2nd-order_ana}), as shown in Fig. 4(d). The
resulting left BRR is also similar to the one in Fig. 3(c). These
confirm our interpretation of the origin of each BRR, in terms of
its born time and returning time.

\begin{figure}
\mbox{\rotatebox{270}{\myscaleboxa{
\includegraphics{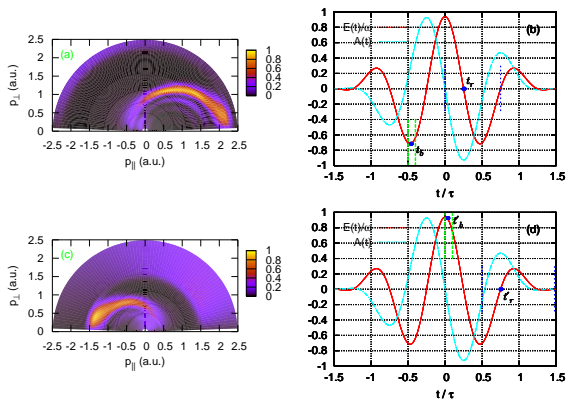}}}}
\caption{(Color online) Analysis of 2D momentum distributions as in
Fig.~3 for the 3-cycle pulse, for the identification of the
tunneling time and rescattering time, for the right- and left-side
BRR electrons. See text.  }
\end{figure}

The position of the BRR in the 2D momentum space can be expressed
as
\begin{eqnarray}
\label{BRR}{\textbf{p}}=-{\textbf{A}}_r+{\textbf{p}}_r
\end{eqnarray}
where ${\textbf{p}}_r$ is momentum vector measured from the center
of the circle which is located at $-{\textbf{A}}_r$ and hence
$p_r$ is the radius of the BRR. The projection of the
photoelectron momentum in the parallel and perpendicular
directions are
\begin{eqnarray}
\label{BRR1}p_{||}&=&p\cos\theta=-A_r-p_r\cos\theta_r \nonumber \\
p_{\perp}&=&p\sin\theta=p_r\sin\theta_r
\end{eqnarray}
where $\theta_r$ is the backscattering angle, ranging from
90$^{\circ}$ to 180$^{\circ}$.

Actually, the BRR in the 2D momentum distribution  is formed by
the cutoff peak in the angle-resolved energy spectra.  To check
the accuracy of the interpretation represented by (\ref{BRR}), we
"measured" the center position and the radius of the right-side
BRR as $|A_r|=0.93$, $p_r=1.18$ in the 2D momentum distributions
shown in Fig 2. We chose $\theta_r=180^{\circ}$, $160^{\circ}$,
$140^{\circ}$, and $120^{\circ}$, which correspond to
$\theta=0^{\circ}$, $11.2^{\circ}$, $22.5^{\circ}$, and
$34^{\circ}$, with the corresponding energies of 2.19, 2.13, 1.94,
and 1.65, respectively. In Fig. 5, we show the angle-resolved
energy spectra at these angles with the predicted electron
energies by arrows. The arrows indeed are located at the peak
positions of the angle-resolved energy spectra.

\begin{figure}
\mbox{\rotatebox{270}{\myscaleboxa{
\includegraphics{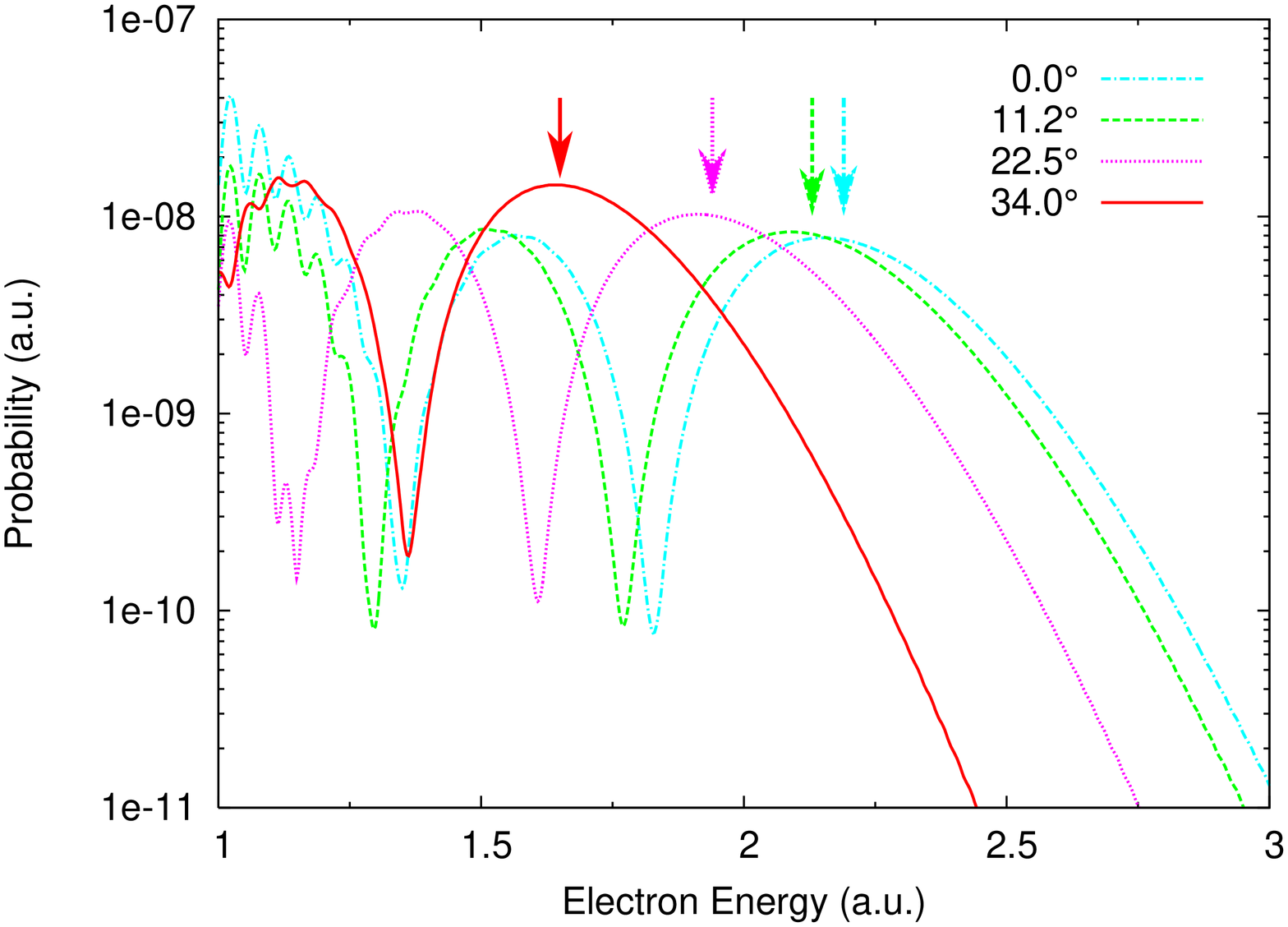}}}}
\caption{(Color online) Angle-resolved energy spectra for atomic
hydrogen by a 5-cycle laser pulse with the wavelength of 800 nm at
the peak intensity of $1.0\times 10^{14}$ W/cm$^{2}$, for four
detector angles of $\theta=0^{\circ}$, $11.2^{\circ}$,
$22.5^{\circ}$, and $34^{\circ}$, respectively. The arrows indicate
the energies predicted by equation (\ref{BRR1}), see text for
detail.}
\end{figure}

\begin{figure}
\mbox{\rotatebox{270}{\myscaleboxb{
\includegraphics{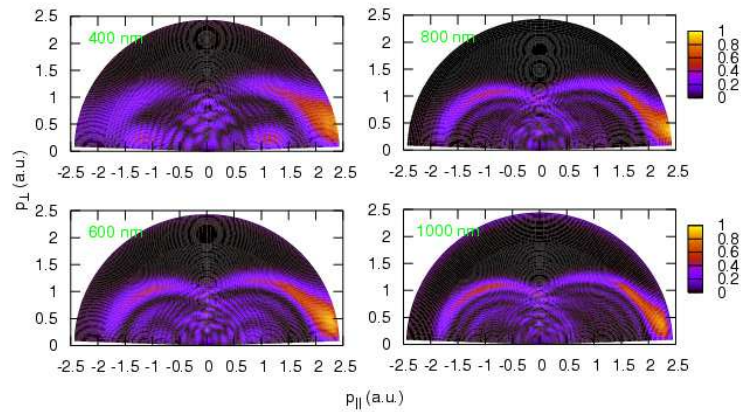}}}}
\caption{(Color online) 2D momentum distributions for atomic
hydrogen by 5-cycle laser pulses with the same Keldysh parameter
$\gamma=1.07$, but for wavelengths of 400, 600, 800, and 1000 nm,
respectively.}
\end{figure}

According to the previous paragraph, the BRR depends on the vector
potential of the pulse only. In Fig. 6, we show the renormalized 2D
momentum spectra for single ionization of hydrogen atom by 5-cycle
laser pulses with fixed Keldysh parameter $\gamma=1.07$. The
wavelengths are 400, 600, 800, and 1000 nm, and the corresponding
peak intensities are $4.0\times 10^{14}$, $1.78\times 10^{14}$,
$1.0\times 10^{14}$, and $6.4\times 10^{13}$ W/cm$^{2}$,
respectively. It can be seen from Fig. 6 that the BRR's exist for
all the cases and the 2D spectra appear to be very similar to each
other. It can also be seen in Fig. 6 that the BRR breaks into
sub-rings and the width of the BRR becomes narrower with increase of
wavelength. These sub-rings are not investigated here.

\subsection{Returning electron wave packets}

By identifying the BRR electrons as due to the backscattering of the
returning electrons by the target ion, the yield along the BRR
should then be proportional to the differential elastic cross
sections of electrons by the target ion. Within the SFA2, the
elastic scattering is treated in the Born approximation, where the
scattering amplitude is proportional to the Fourier transform of the
ion potential
\begin{eqnarray}
\label{formfactor}V(\textbf{q})=-\frac{1}{4\pi} \int
\exp(i\mathbf{q}\cdot \textbf{r}) V(\textbf{r})d\textbf{r},
\end{eqnarray}
here $\textbf{q}$ is the momentum transfer which is related to the
rescattering angle $\theta_r$ and the radius $p_r$ of the BRR by
\begin{eqnarray}
\label{momentum_transfer}q=2p_r \sin(\theta_r/2).
\end{eqnarray}

\begin{figure}
\mbox{\rotatebox{270}{\myscaleboxa{
\includegraphics{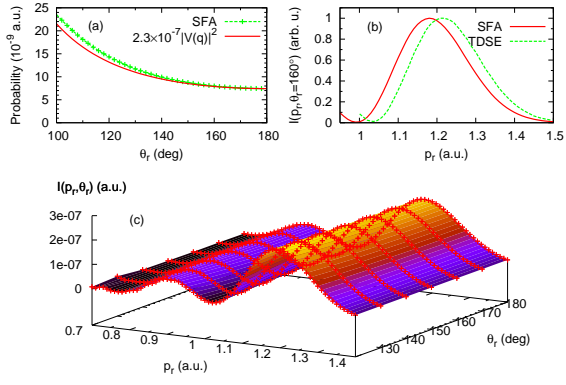}}}}
\caption{(Color online) Right-side BRR for atomic hydrogen by a
5-cycle laser pulse at the peak intensity of $1.0\times 10^{14}$
W/cm$^{2}$ with the wavelength of 800 nm. (a) Comparison of the
first Born elastic scattering cross sections with the  angular
distributions on BRR calculated from SFA2; (b) Comparison of the
electron wave packet from SFA2 with that from TDSE at
$\theta_r=160^{\circ}$; (c) The wave packets of the returning
electrons are shown to be independent of the scattering angles.}
\end{figure}

Fig. 7(a) shows the comparison of the first Born electron elastic
scattering cross sections with the angular distributions on the
right-side BRR in the 2D momentum spectra shown in Fig. 2(a). For
the comparison presented in Fig. 7(a), the radius of BRR is taken to
be $p_r=1.2$. It can be seen that for $\theta_r>100^{\circ}$, the
differential elastic scattering cross sections are very close to the
photoelectron angular distributions along BRR, indicating that
elastic scattering of electrons by the parent ion can be extracted
from the angular distribution along the BRR.

Expanding the rescattering idea further, we then ask if it is
possible to treat the electron yield on the BRR as due to the
backscattering of a returning electron wave packet. To test this
idea, we write
\begin{eqnarray}
\label{key_relation1}\frac{\partial^2P}{\partial E
\partial \hat{\textbf{p}}}|_{\varphi=\text {cons}}=I(p_r,\theta_r)|V(\textbf{q})|^2.
\end{eqnarray}
On the left side in (\ref{key_relation1}), the angular distribution
is obtained on any plane containing the $z$ axis in the momentum
space due to the cylindrical symmetry. Note that $|V(\textbf{q})|^2$
is the elastic differential cross section for each ion by an
electron with energy $E_r=p_r^2/2$ (see (\ref{momentum_transfer}))
while the actual energy of electron in the laser field is
$E=[-{\textbf{A}}_r+{\textbf{p}}_r]^2/2$. In Fig. 7(b), we show the
extracted $I(p_r,\theta_r)$ at the angle of $\theta_r=160^{\circ}$
from the right-side BRR in Fig. 2. The wave packet extracted from
SFA2 is essentially identical to that from the TDSE except for a
small shift of the center from 1.18 to 1.21. Here, we see that the
radius of BRR is determined by the center of the wave packet.

\begin{figure}
\mbox{\rotatebox{270}{\myscaleboxa{
\includegraphics{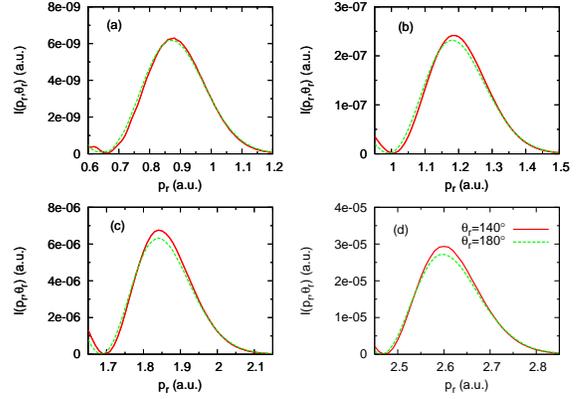}}}}
\caption{(Color online) Distributions of the momentum of the
right-side BRR electrons for atomic hydrogen by 5-cycle laser pulses
with the wavelength of 800 nm at the peak intensities of (a)
$5.0\times 10^{13}$, (b) $1.0\times 10^{14}$, (c) $2.5\times
10^{14}$, and (d) $5.0\times 10^{14}$ W/cm$^{2}$. }
\end{figure}

According to the classical or semiclassical theory the maximal
energy of the returning electron is $3.17U_p$. Extending this idea
for each optical cycle, we then have $E_r=p_r^2/2=3.17\bar{U}_p$,
where $\bar{U}_p=A_r^2/4$. Consequently, we get
\begin{eqnarray}
\label{key_relation2}p_r=1.26A_r.
\end{eqnarray}
For a 5-cycle laser pulse at the peak intensity of $1.0\times
10^{14}$ W/cm$^{2}$ with the wavelength of 800 nm, $A_r=0.93$ for
the right-side BRR, corresponding to $p_r=1.17$, which agrees with
 the SFA2 calculation.

The wave packet extracted from (\ref{key_relation1}) should not
depend on $\theta_r$ if the electron on the BRR comes entirely
from the backscattering   when the scattering angle is large. We
found this is the case for $\theta_r>120^{\circ}$, as plotted in
Fig. 7(c).

We next check the relation (\ref{key_relation2}) for different
intensities. In Fig. 8 are presented the wave packets extracted
from the right-side BRR for the photoelectrons generated by
5-cycle, 800nm laser pulses for peak intensities of 0.5, 1.0, 2,5
and $5.0\times 10^{14}$ W/cm$^{2}$. First we note that the wave
packets derived from the two scattering angles are nearly the
same. For these pulses, the vector potentials at the return time
have values $A_r=0.65$, 0.93, 1.46 and 2.08, respectively. From
(\ref{key_relation2}), the radii are calculated to be $p_r=0.82$,
1.17, 1.84 and 2.62, which   compare  well with the peak positions
of 0.86, 1.18, 1.84 and 2.60, from the extracted electron wave
packets shown in Fig. 8.

\subsection{Returning electron wave packet for long pulses}

The analysis of BRR so far has been limited to short pulses. For
longer pulses, the BRR electrons on each side can be generated
from several born times, each separated from the previous one by
one full optical cycle. Interference from these coherent electron
bursts results in  the characteristic  ATI peaks.

In Fig. 9, we show the SFA2 results for 10- and 20-cycle pulses at
the peak intensity of $1.0\times 10^{14}$ W/cm$^{2}$ and
wavelength of 800 nm.  For these "long" pulses, accurate TDSE
results are difficult to obtain. Figs. 9(a) and (b) show that the
electron yields on the BRR are oscillatory, but the envelope is
well reproduced by the differential elastic scattering cross
sections. The oscillations are attributed to the electron wave
packets, see Figs. 9(c) and (d). Note that the envelopes of the
electron wave packets in these two figures are similar to the
smooth wave packet for the 5-cycle pulse shown in Fig. 7(b). The
peak positions of the envelopes for 10- and 20-cycle pulses are at
1.16 and 1.17, respectively from the figure, as compared to 1.17
from Fig. 7(b).

\begin{figure}
\mbox{\rotatebox{270}{\myscaleboxa{
\includegraphics{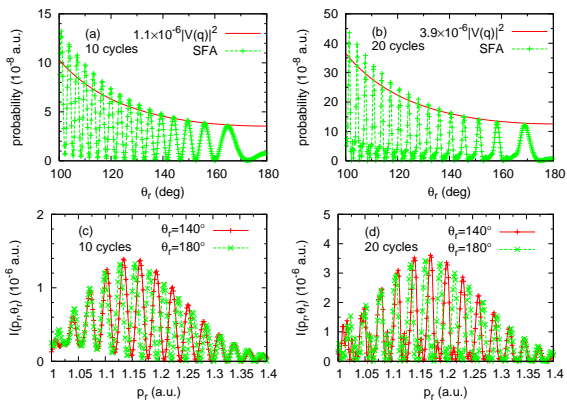}}}}
\caption{Ionization of hydrogen by laser pulses at the peak
intensity of $1.0\times 10^{14}$ W/cm$^{2}$ with the wavelength of
800 nm and the full durations of 10 and 20 cycles. (a,b): Comparison
of the first Born elastic scattering cross sections with the SFA2
angular distributions of the BRR electrons; (c,d): Distributions of
$I(p_r,\theta_r)$ at $\theta_r=140^{\circ}$ and $180^{\circ}$.}
\end{figure}

For long pulses, electrons which have been released earlier may
return at different times. So far we have considered the dominant
first return. For later returns it is generally believed that the
probabilities would be smaller since the electron wave packet
expected to spread in time.    Based on the SFA2, we can  estimate
the wave packets for the first and the later returns. For this
purpose, we consider a 25-cycle pulse. We assume   that the
electrons are born in the interval $[-0.5\tau, -0.4\tau]$. For the
first return we isolate backscattering occurring within the time
interval $[0.0, 0.5\tau]$. For the second and the third returns, the
intervals are chose to be $[0.5\tau, 1.0\tau]$ and $[1.0\tau,
1.5\tau]$, respectively, i.e., each is half an optical cycle later
from the previous return. From the calculated 2D spectra for each
return, we extracted the electron wave packets, see Fig. 10. It is
clear that the wave packet from the first return has the highest
yield and highest momentum. The wave packet from second return has
lower momentum than from the third return, but the peak intensity is
about twice higher. The wave packets "derived" from this model are
qualitatively similar to those obtained from the classical
simulation by Tong \emph{et al} \cite{tong93pra}, but with larger
width and smaller strength since spreading of the wave packet is
included in the SFA2 but not in the classical simulation of the
latter.

\begin{figure}
\mbox{\rotatebox{270}{\myscaleboxa{
\includegraphics{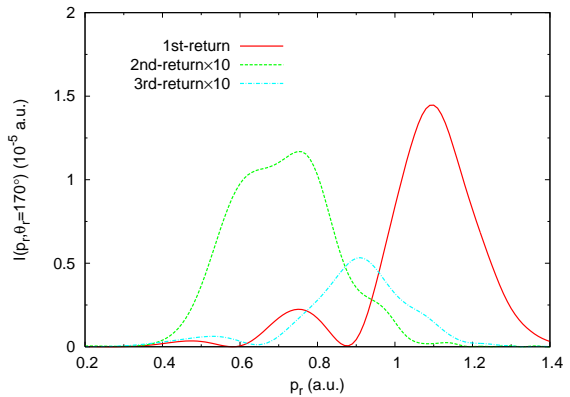}}}}
\caption{Distributions of $I(p_r,\theta_r)$ of the BRR electrons,
born at the time ($-0.5\tau$, $-0.4\tau$) and returning at the
1st-return time (0.0, $0.5\tau$), the 2nd-return time ($0.5\tau$,
$1.0\tau$) and the 3rd-return time ($1.0\tau$, $1.5\tau$), at
$\theta_r=170^{\circ}$ for ionization of hydrogen by a 25-cycle
laser pulse at the peak intensity of $1.0\times 10^{14}$ W/cm$^{2}$
with the wavelength of 800 nm.}
\end{figure}

\section{Summary}

In this paper we studied the two-dimensional momentum spectra of
high-energy photoelectrons of a hydrogen atom in an intense laser
pulse. We focussed on electrons that are on the back rescattered
ridges (BRR). These electrons have been identified initially from
solving the time-dependent Schr\"{o}dinger equation (TDSE) and they
were interpreted as due to the backscattering of the returning
electrons by the target ion. Using the second order strong field
approximation (SFA2) where the rescattering is accounted for to the
first order, we showed that BRR electrons also appear in the SFA2
calculation. By analyzing the results from SFA2 we have been able to
identify the time of birth of the electrons and the return time
where these electrons are backscattered by the target ion. From the
yields of the electrons along the BRR, we further showed that it is
possible to extract the returning electron wave packet as well as
the elastic scattering cross sections between the returning
electrons and the target ion. The electron wave packets extracted
from SFA2 have been shown to be very close to those extracted from
solving the TDSE. Since the SFA2 calculation is much simpler, this
allows us to analyze the returning electron wave packets
conveniently for laser pulses of different intensities and
durations. We comment that the yield along the BRR calculated using
the SFA2 does not predict the correct elastic scattering cross
section between the electron and the target ion since it was based
on the first order theory. For the present atomic hydrogen target,
on the other hand, the elastic scattering cross sections calculated
from the first-order theory and the exact results are identical -- a
special property of the Coulomb potential. While results have been
presented here only for atomic hydrogen target, similar comparison
with the same conclusion has been made for other atoms. In addition,
there is no reason to expect that the same conclusion would not hold
for molecular targets where accurate TDSE solutions at the same
level of accuracy as in atoms  is computationally not feasible.

\section{Acknowledgment}

This work was supported in part by Chemical Sciences, Geosciences
and Biosciences Division, Office of Basic Energy Sciences, Office of
Science, US Department of Energy. TM is also supported by financial
aids from the research fund of the University of
Electro-communications,the 21st century COE program on "Coherent
Optical Science" and the Ministry of Education, Culture, Sports,
Science, and Technology, Japan.

\end{document}